\documentclass[final]{article}
\usepackage{color}
\usepackage{amssymb,amsmath,amsthm,latexsym,graphicx}
\usepackage[color]{showkeys}
\definecolor{refkey}{gray}{.75}
\usepackage{hyperref}
\usepackage{booktabs}
\definecolor{refkey}{gray}{.75}
\usepackage{rotating}
\usepackage{rotfloat}

\newcommand{\ra}[1]{\renewcommand{\arraystretch}{#1}}

\def\p{\partial}
\def\tilde{\widetilde}

\def\bs{\boldsymbol}

\def\ub{\mathbf{u}}

\def\kb{\bs{\mathrm{k}}}
\def\hbar{\overline{h}}

\def\kg{\textrm{kg}}
\def\m{\textrm{m}}

\def\C{\mathcal{C}}

\def\lop{{\small{\mathcal{L}^2(\Omega; P)}}}

\newtheorem{assumption}{Assumption}[section]

\begin{document}
\title{The multi-level Monte Carlo method for simulations of turbulent
  flows}

\author{ Qingshan Chen\\
  Department of Mathematical Sciences\\
  Clemson University \\
  Clemson, SC 29631, USA\\
  \and Ju Ming\thanks{Corresponding author: jming@csrc.ac.cn}\\
Beijing Computational Science Research Center\\
Haidian District, Beijing 100193, CHINA}

\maketitle

\abstract{In this paper the application of the multi-level Monte Carlo (MLMC)
  method on numerical simulations of turbulent flows with
  uncertain parameters is investigated. Several strategies for
  setting up the MLMC method are presented, and
  the advantages and disadvantages of
  each strategy are also discussed. A numerical experiment is carried
  out using the
  Antarctic Circumpolar Current (ACC) with uncertain,
  small-scale bottom
  topographic features. It is demonstrated that,
  unlike the pointwise solutions, the averaged volume transports
  are correlated across grid resolutions,  and the MLMC method could increase simulation efficiency without losing accuracy in uncertainty assessment.  }

\section{Introduction}

Monter Carlo (MC) method has long been known to mathematicians and physicians  as one of the most versatile and widely used computational algorithms. With the advantage of a dimension-independent convergence rate, it is regarded as the most efficient method to overcome the \emph{the curse of dimensionality} (\cite{Metropolis:2012ft}). However, the slow convergent rate, $\mathcal{O}({N}^{-1/2})$(where $N$ denotes the sample size), often results in unaffordable computational cost to generate high-resolution samples with a large sample size. Specifically, when MC is applied to the complex system models described by differential equations with uncertainties, which usually arise from e.g., data inaccuracies and information loss, the computational cost will dramatically (polynomially) grow due to the larger sample size required as one moves onto high-resolution meshes. To mitigate such growth, many efforts such as quasi-Monte Carlo method (\cite{2013-paper-Sloan-QMC, 2012-paper-Kuo-QMC, 1993-paper-Niederreiter-QMC}), variance reduction method (\cite{1964-book-Hammersley-VR}), importance sampling and stratified sampling method (\cite{2015-paper-Bugallo-IS, 2001-book-Liu-MC}), etc, have been made to speed up the convergence of MC. 

Besides these ameliorated methods, the mulit-level Monte Carlo (MLMC) method has attracted much attention for its promising potential in the reduction of computational complexity of uncertainty quantification (UQ) problems (e.g., \cite{Heinrich:2001jl, {Kebaier:2005ce},{Giles:2008gc}} and references therein). 
Similar to multi-grid method for iteratively solving large linear deterministic systems (\cite{Wesseling:1992ug}), the MLMC algorithm utilizes a hierarchy of resolutions instead of one. The basic idea, roughly speaking, is to obtain
independent numerical samples on the coarse grids (higher level),
then improve the results on the fine grids (lower level) iteratively.
The variance decays with level at a faster rate than the computational expense increases.
It can be shown that MLMC could strike a balance between the efficiency and accuracy in solving
the UQ problems and obtaining the quantity of interests (QoI).

There is a large body of literature on MLMC,
and some relevant references are listed as follows: Barth et al
(\cite{Barth:2011gz}) couples the MLMC method with the finite element
method (FEM) to solve stochastic elliptic equations, and presents rigorous error analysis. Mishra et al (\cite{Mishra:2012ec, Mishra:2012ea,
  Mishra:2012cv, {Mishra:2016hv}}) couple the MLMC method with the
finite volume method for hyperbolic systems. Kornhuber et al
(\cite{Kornhuber:2014iz})  applies the MLMC with FEM to study
stochastic elliptic variational inequalities. Li et al
(\cite{Li:2016ie}) couples the MLMC with the weak Galerkin method to
study the elliptic equations. For a survey of the MLMC and the
literature on its applications, see \cite{Giles:2015dd}.

In this paper we are concerned with the applicability of the MLMC method for
long-term simulations of turbulent geophysical flows with uncertain parameters.
For turbulent flows, pointwise behaviors of the solutions are no longer
relevant. In fact, after the initial spinup period, the difference between solutions
on two different meshes is spatially uncorrelated,
even if all the other settings are the same. 
Thus, the usual notion of error convergence, e.g.,
the pointwise error estimates under certain norms, no longer applies.
Due to this unreliable nature of the pointwise solutions, 
the research  objective  of turbulence simulations is
often focused  on computing certain aggregated QoI's,
such as the global mean of sea surface temperature,
instead of the pointwise solutions. 
Our main motivation for this work is to adapt the analysis of MLMC to QoI under some
verifiable assumptions. Both the assumptions and the conclusions will
be examined using the Antarctic Circumpolar Current
(ACC) model.

Turbulence models often include closures to account for unresolved eddy
activities. These closures in general need to be adjusted according to the
level of mesh resolutions. This is a dramatic departure
from the situation involving steady-state or laminar flows, where the
discrete model is kept the same, and only grid resolutions vary. But
this departure does not automatically invalidate the MLMC for turbulent
flows. Eddy closures are implemented to prevent instability and to
improve qualitative large-scale behaviors of the solution. However, the
accuracy of the estimate of the QoI is
aligned with the grid resolutions, i.e.,  the estimate will
improve or worsen as the mesh refines or
coarsens.
Based on this premise, the effectiveness and applicability of
MLMC could be expected for simulations of turbulent flows.

The numerical scheme used in this paper is a staggered C-grid
finite difference finite volume scheme (\cite{Ringler:2010io})  based on a Voronoi tessellation (VT, \cite{Du:1999gs, Du:2003ka}). The
VT primarily consists of pentagons and hexagons, and thus nesting
between different levels of meshes is impossible, which implies a
direct comparison between the solutions on two different meshes is also
impossible.  This would result in a major hurdle in applying MLMC to steady-state or laminar flows,  but for  turbulent flows, the pointwise
behaviors of the solution are uncorrelated, and the focus is instead on
QoI. Thus, the issue with mesh matching is irrelevant
here.

The objective of the present paper is two-fold: {\bf(i).} to
explore the effectiveness of the MLMC method in the presence of the
challenges associated with turbulent flows. {\bf(ii).} to explore
an optimal way to set up the MLMC simulations. The rest
of the paper is organized as follows. In Section
\ref{sec:monte-carlo-multi}, we briefly review the MC and the MLMC
methods, and detail the possible strategies for setting up the MLMC
simulations. In Section \ref{sec:numer-exper-using}, we apply the MLMC
method to a turbulent channel flow mimicking the ACC, and examine the effectiveness of the method under various
strategies. The paper ends with some concluding remarks in Section
\ref{sec:discussions}.

\section{The Monte Carlo and the multi-level Monte Carlo
  methods}\label{sec:monte-carlo-multi}
We designate the QoI  to be calculated by $U$, which
can be e.g., volume transport, mean sea-surface temperature (SST), etc.

We denote the number of levels of grid resolution by $L$, and the
resolution at each level by $r_l$, $1\leq l\leq L$, and the highest
resolution by $r\equiv r_1$. We assume that
\begin{equation}
  \label{eq:1}
  r_l = 2r_{l-1} = 2^{l-1}r_1 \equiv 2^{l-1} r.
\end{equation}

\begin{assumption}
We assume that, at each level, the computational cost is proportional
to the total number of spatial-temporal degrees of freedom $N_l$. For
simplicity, in the sequel, we identify the computational cost with
$N_l$. We further assume that the total number of degrees of freedom
is proportional to $r_l^{-3}$, that is,
\begin{equation}
  \label{eq:2}
  N_l = C_1 r_l^{-3}.
\end{equation}
\end{assumption}

We designate the total number of degrees of freedom at the highest
resolution $r_1 \equiv r$ by
\begin{equation}
  \label{eq:3}
  N \equiv N_1 = C_1 r^{-3},
\end{equation}
where $C_1$ is a constant. The cubic relation between $N$ and $r$ is tailored towards models of large-scale geophysical
flows, where the vertical resolution is often held fixed and the time
step size varies linearly according to the horizontal resolution.
From (\ref{eq:1}) and (\ref{eq:2}) it is derived that
\begin{equation}
  \label{eq:4}
  \dfrac{N_l}{N} = 8^{-(l-1)}.
\end{equation}

\subsection{The Monte Carlo method}\label{sec:monte-carlo-method}
We recall the classical MC method as it is applied to the
ensemble simulations at a fixed resolution $r$. The numerical
approximation of the QoI $U$ at this resolution is
denoted by $U_r$, and the computational cost of each individual
simulation by $N$, which is related to the grid resolution through
(\ref{eq:3}). We denote each realization by a superscript $m$, as in
$U^m$ and $U_r^m$, $1\leq m\leq M$. The MC mean is defined as
\begin{equation}
  \label{eq:5}
  \mbox{\rm\sffamily E}_M\left[ U_r\right]: = \dfrac{1}{M} \sum_{m=1}^M U_r^m.
\end{equation}

We now examine the difference between the sample mean and the
expectation $\mbox{\rm\sffamily E}[U]$ of the true solution $U$. We use the standard notations for the $\sigma$-finite probability space $(\Omega,\mathcal{F},P)$, where sample space $\Omega$  is a set of all possible outcomes, $\mathcal{F}$ is a $\sigma$-algebra of events, and $P : \mathcal{F}\rightarrow[0,1]$ is a probability measure.

\begin{equation}
  \label{eq:6}
  \left\| \mbox{\rm\sffamily E}[U] - \mbox{\rm\sffamily E}_M[U_r]\right\|_\lop \leq
\|\mbox{\rm\sffamily E}[U] - \mbox{\rm\sffamily E}_M[U]\|_\lop + \|\mbox{\rm\sffamily E}_M[U] - \mbox{\rm\sffamily E}_M[U_r]\|_\lop.
\end{equation}
We note that, by the Central Limit Theorem,
\begin{equation}
  \label{eq:7}
  \|\mbox{\rm\sffamily E}[U] - \mbox{\rm\sffamily E}_M[U]\|_\lop = \dfrac{\delta}{\sqrt{M}},
\end{equation}
where $\delta$ is the standard deviation in the true solution. For the
second term on the right-hand side of (\ref{eq:6}),
\begin{align*}
&  \|\mbox{\rm\sffamily E}_M[U] - \mbox{\rm\sffamily E}_M[U_r]\|_\lop= \|\mbox{\rm\sffamily E}_M[U-U_r]\|_\lop
= \|\dfrac{1}{M}\sum_{m=1}^M(U^m - U_r^m)\|_\lop \\ \leq & \dfrac{1}{M}\sum_{m=1}^M \left\|U^m - U^m_r\right\|_\lop= \mbox{\rm\sffamily E}_M\left[\|U-U_r\|_\lop\right]=\left\|U-U_r\right\|_\lop.
\end{align*}

\begin{assumption}
We assume that the $L^2$-norm of the error in the quantity of interest
is proportional to $r^\alpha$, where $\alpha$ designates the rate of
convergence regarding the quantity.
\end{assumption}
That is, designating the $L^2$-norm of the error at the resolution $r$
by $e$, we may write that
\begin{equation}
  \label{eq:64}
e \equiv \| U- U_r\|_\lop = C_2r^\alpha,
\end{equation}
where $C_2$ is a constant independent of the grid  resolution.
Hence, concerning the MC mean of the true solution and the MC
mean of the approximate solution, we have
\begin{equation}
  \label{eq:8}
 \|\mbox{\rm\sffamily E}_M[U] - \mbox{\rm\sffamily E}_M[U_r]\|_\lop \leq e
= C_2 r^\alpha.
\end{equation}
Combining (\ref{eq:7}) and (\ref{eq:8}) yields
\begin{equation}
  \label{eq:9}
\|\mbox{\rm\sffamily E}[U] - \mbox{\rm\sffamily E}_M[U_r]\|_\lop \leq
\dfrac{\delta}{\sqrt{M}} + e.
\end{equation}
The first term on the right-hand side represents the discretization
error of the probability space, and the second term represents the
discretization error of the temporal-spatial space.

For a given resolution $r$, the the spatial-temporal discretization
error is fixed. The sample size $M$ should be chosen so that the
probability space discretization error is on the same order as the
temporal-spatial discretization error. Thus, we set
\begin{align}
  \dfrac{\delta}{\sqrt{M}} &= e,\   M = \dfrac{\delta^2}{e^2}\label{eq:10}.
\end{align}
Given this choice of $M$, now the combined errors in the MC mean can
be given,
\begin{equation}
  \label{eq:11}
\|\mbox{\rm\sffamily E}[U] - \mbox{\rm\sffamily E}_M[U_r]\|_\lop
\leq 2 e.
\end{equation}
The total computational cost for the Monte Carlo method, $\mathcal{C}_{\mathrm{\tiny MC}}$,  can also be calculated, using
\eqref{eq:3}, \eqref{eq:64}, and \eqref{eq:10},
\begin{equation}
  \label{eq:12}
  \mathcal{C}_{\mathrm{\tiny MC}} = N\cdot M =
  \dfrac{\delta^2}{C_1^{\frac{2}{3}\alpha} C^2_2} N^{1+\frac{2}{3}\alpha}.
\end{equation}
The total computational cost for the ensemble simulation using the
conventional MC method grows polynomially in terms of the
computational cost for each individual simulation, and the degree of
the polynomial is $1+\frac{2}{3}\alpha$. Also as expected, larger
deviation $\delta$ in the true solution would demand more
computational resource.

\subsection{The multi-level Monte Carlo method}
We denote the numerical approximation of $U$ at each level by $U_l$,
$1\le l\le L$, and each realization of $U_l$ by $U^m_l$, $1\le m\le
M_l$.

We note that the numerical approximation $U_1$ at the lowest level
(highest resolution) can be decomposed as
\begin{equation}
  \label{eq:13}
  U_1 = \sum_{l=1}^{L-1}(U_l - U_{l+1}) + U_L.
\end{equation}
Then clearly,
\begin{equation}
  \label{eq:14}
  \mbox{\rm\sffamily E}[U_1] = \sum_{l=1}^{L-1}\mbox{\rm\sffamily E}[U_l - U_{l+1}] + \mbox{\rm\sffamily E}[U_L].
\end{equation}
In practice, the mean is approximated by MC mean, and as it has
been shown above, the accuracy of such approximation is determined by
two competing factors, the variance in the random variable $\delta$ and the
sample size $M$. A larger variance requires a larger sample size. The
success of the MLMC method is built on the hypothesis that the
variance of the difference between two solutions at successive levels
is much smaller than the variance of each individual solution, and
thus requires a much smaller sample size. We now define the $L$-level
sample mean of $U$, $ \mbox{\rm\sffamily E}^L[U]$, as
\begin{equation}
  \label{eq:15}
     \mbox{\rm\sffamily E}^L[U] = \sum_{l=1}^{L-1} \mbox{\rm\sffamily E}_{M_l}[U_l - U_{l+1}] +  \mbox{\rm\sffamily E}_{M_L}[U_L],
\end{equation}
where $M_l$ represents the sample size, and the sample mean $ \mbox{\rm\sffamily E}_{M_l}$
at each level is defined in the same way as (\ref{eq:5}).
The relation between $M_l$ and the total sample size $\tilde M_l$ at
each level is as follows,
\begin{equation}
  \label{eq:16}
  \left\{
    \begin{aligned}
      &\tilde M_1 = M_1,\\
      &\tilde M_l = M_{l-1} + M_l,\qquad 2\le l\le L.
    \end{aligned}\right.
\end{equation}

We now examine the theoretical mean $\mbox{\rm\sffamily E}[U]$ and the $L$-level sample
mean $\mbox{\rm\sffamily E}^L[U]$.
\begin{equation}
  \label{eq:17}
  \| \mbox{\rm\sffamily E}[U] -  \mbox{\rm\sffamily E}^L[U]\|_\lop \leq \| \mbox{\rm\sffamily E}[U] -  \mbox{\rm\sffamily E}[U_1]\|_\lop + \| \mbox{\rm\sffamily E}[U_1] -  \mbox{\rm\sffamily E}^L[U]\|_\lop.
\end{equation}
We note that,
\begin{align*}
  \|\mbox{\rm\sffamily E}[U] - \mbox{\rm\sffamily E}[U_1]\|_\lop = \|\mbox{\rm\sffamily E}[U-U_1]\|_\lop \leq & \mbox{\rm\sffamily E}[\|U-U_1\|_\lop] \\ = & \|U-U_1\|_\lop.
\end{align*}
By the standing assumption \eqref{eq:64}, we obtain an estimate of the
first term on the right-hand side of \eqref{eq:17},
\begin{equation}
  \label{eq:18}
\|\mbox{\rm\sffamily E}[U] - \mbox{\rm\sffamily E}[U_1]\|_\lop   \leq e\equiv  C_2\cdot r_1^\alpha.
\end{equation}
For the second term on the right-hand side of \eqref{eq:17}, using the
relation (\ref{eq:14}) and the definition (\ref{eq:15}), we find that
\begin{align*}
  &\|\mbox{\rm\sffamily E}[U_1] - \mbox{\rm\sffamily E}^L[U]\|_\lop \\
  =& \left\| \sum_{l=1}^{L-1}\left( \mbox{\rm\sffamily E}[U_l - U_{l+1}] -
    \mbox{\rm\sffamily E}_{M_l}[U_l-U_{l+1}]\right)  + \mbox{\rm\sffamily E}[U_L] - \mbox{\rm\sffamily E}_{M_L}[U_L] \right\|_\lop\\
  \leq & \sum_{l=1}^{L-1}\left\| \mbox{\rm\sffamily E}[U_l - U_{l+1}] -
   \mbox{\rm\sffamily E}_{M_l}[U_l-U_{l+1}] \right\|_\lop  + \|\mbox{\rm\sffamily E}[U_L] - \mbox{\rm\sffamily E}_{M_L}[U_L] \|_\lop\\
  \leq & \sum_{l=1}^{L-1} \dfrac{\delta[U_l - U_{l+1}]}{\sqrt{M_l}} +
  \dfrac{\delta[U_L]}{\sqrt{M_L}}.
\end{align*}
For $1\leq l\leq L-1$, by the standard definition of variance, we
deduce that
\begin{align*}
  &  \delta[U_l - U_{l+1}] = \mbox{\rm\sffamily E}[|U_l - U_{l+1}|^2] - \mbox{\rm\sffamily E}[U_l - U_{l+1}]^2\\
  \leq & \mbox{\rm\sffamily E}[|U_l - U_{l+1}|^2]  \leq  2\left( \mbox{\rm\sffamily E}[|U_l - U|^2] + \mbox{\rm\sffamily E}[|U_{l+1} - U|^2]\right),
\end{align*}
and, again, by
the standing assumption \eqref{eq:64},
\begin{equation}
\delta[U_l - U_{l+1}]
  \leq  2C_2^2(1+4^\alpha)\cdot r_l^{2\alpha}.\label{eq:49}
\end{equation}
For the variance at the lowest resolution, $\delta[U_L]$, we again
start from the definition,
\begin{align*}
  \delta[U_L] &= \|U_L-\mbox{\rm\sffamily E}[U_L]\|_\lop
  = \|U_L - U + U-\mbox{\rm\sffamily E}[U] + \mbox{\rm\sffamily E}[U] - \mbox{\rm\sffamily E}[U_L]\|_\lop\\
  &\leq \|U_L-U\|_\lop + \|U-\mbox{\rm\sffamily E}[U]\|_\lop + \|\mbox{\rm\sffamily E}[U-U_L]\|_\lop\\
  &\leq 2\|U-U_L\|_\lop + \|U-\mbox{\rm\sffamily E}[U]\|_\lop \leq 2C_2r_L^\alpha + \delta[U].
\end{align*}
Combining the last three estimates, we obtain
\begin{align*}
  \|\mbox{\rm\sffamily E}[U_1] - \mbox{\rm\sffamily E}^L[U]\|_\lop  &\leq C_2\sqrt{2(1+4^\alpha)}
  \sum_{l=1}^{L-1}\dfrac{r_l^\alpha}{\sqrt{M_l}} +
    \dfrac{2C_2r_L^\alpha}{\sqrt{M_L}} +
    \dfrac{\delta[U]}{\sqrt{M_L}}.
\end{align*}
Assuming that $\alpha\ge 0$, which should be true for all practically
useful numerical
schemes, we may bring the second term on the right-hand side into the
summation, and we thus obtain
\begin{equation}
\label{eq:19}
\|\mbox{\rm\sffamily E}[U_1] - \mbox{\rm\sffamily E}^L[U]\|_\lop
\leq
   C_2\sqrt{2(1+4^\alpha)}\sum_{l=1}^L\dfrac{r_L^\alpha}{\sqrt{M_l}} + \dfrac{\delta[U]}{\sqrt{M_L}}.
\end{equation}
Combining \eqref{eq:17}, \eqref{eq:18} and \eqref{eq:19} leads us to
\begin{equation}
  \label{eq:20}
  \|\mbox{\rm\sffamily E}[U] - \mbox{\rm\sffamily E}^L[U]\|_\lop \leq e +
  C_2\sqrt{2(1+4^\alpha)}\sum_{l=1}^L\dfrac{r_l^\alpha}{\sqrt{M_l}} +
  \dfrac{\delta[U]}{\sqrt{M_L}}.
\end{equation}

This estimate shows that the error in the
$L$-level sample mean of the quantity $U$ can be attributed to three
components: the temporal-spatial discretization error (first term), the probability
space discretization error (third term), and the error for using a
multi-level structure (the second term).
So far, the sample size at each level, $M_l$ has been left
to be determined. Determining the sample size will be a delicate balancing
act between controlling the computational cost and controlling the
error. The potential of the MLMC method lies in the
fact that, within the probability space discretization
error (the third term), the standard deviation of the analytical
solution is divided by
the sample size at the highest level (lowest resolution), where the
computational cost for an individual simulation is the
lowest. We should also note that the first term, the temporal-spatial
discretization error, is not affected by the sample size at any level.
Hence, a general principle for determining the sample size is to make
sure the probability space discretization error (the third term) and
each term in the
summation (the second term) is roughly on the order of the
temporal-spatial
discretization error (the first term), or smaller.
By the this principle, we know exactly what the sample size at the
lowest resolution should be (see \eqref{eq:10}). But to reach this
sample size starting from the highest resolution can take many
different paths. Here, we explore several different strategies for
determining the sample size at each level.
\\

\noindent{\itshape Strategy \#1}\\
Our first strategy is to choose the sample size for each level  so that
each term in the summation of \eqref{eq:20} is equal or smaller than the temporal-spatial
discretization error. Hence we set
\begin{equation*}
 e \equiv  C_2r_1^\alpha =
  C_2\sqrt{2(1+4^\alpha)}\dfrac{r_l^\alpha}{\sqrt{M_l}},
\end{equation*}
which leads to
\begin{equation}
  \label{eq:21}
  M_l = 2(1+4^\alpha)\cdot 2^{2\alpha(l-1)},\qquad 1\le l\le L.
\end{equation}
The number of $L$ is determined by requiring that the sample size at
the lowest resolution, $M_L$ be sufficiently large to make the
probability discretization error be on the same order as the
temporal-spatial discretization error, that is,
\begin{equation}
  \label{eq:22}
\dfrac{\delta[U]}{\sqrt{M_L}} = e,
\end{equation}
from which, and \eqref{eq:21}, we deduce that
\begin{equation}
  \label{eq:23}
  L = 1 + \dfrac{2\log \delta[U] - 2\log e - \log
    2(1+4^\alpha)}{2\alpha\log 2},
\end{equation}
or, using \eqref{eq:18},
\begin{equation}
\label{eq:54}
  L = \dfrac{\log\delta[U]}{\alpha\log 2} - \dfrac{\log r_1}{\log
    2} +1 - \dfrac{2\log C_2 + \log 2(1+4^\alpha)}{2\alpha\log 2}.
\end{equation}
We note from \eqref{eq:3} that
\begin{equation*}
  \log r_1 = \dfrac{\log C_1 - \log N}{3}.
\end{equation*}
Substituting this expression into \eqref{eq:54} yields
\begin{equation}
\label{eq:55}
  L = \dfrac{\log\delta[U]}{\alpha\log 2} + \dfrac{\log N}{3\log
    2} +1 - \dfrac{\log C_1}{3\log 2} - \dfrac{2\log C_2 + \log
    2(1+4^\alpha)}{2\alpha\log 2}.
\end{equation}
The expression on the right-hand size indicates that, generally, a
larger variance in the analytical solution requires more levels. Under
the same order of convergence ($\alpha$), and the same constant
coefficients $C_1$ and $C_2$, a higer number of degrees of freedom ($N$,
or, in other words, a finer mesh) also requires more levels.

Based on this strategy, the total error in the $L$-level sample mean
is
\begin{equation}
\label{eq:24}
  \|\mbox{\rm\sffamily E}[U] - \mbox{\rm\sffamily E}^L[U]\|_\lop \leq (L+2) e.
\end{equation}

We denote the computational cost under this strategy as
$\mathcal{C}_{MLMC1}$, which can be calculated as
\begin{align*}
  \C_{MLMC1} &= \sum_{l=1}^L \tilde M_l N_l = \sum_{l=1}^{L-1}
  M_l(N_l+N_{l+1}) + M_L N_L\\
  &= \sum_{l=1}^{L-1} 2(1+4^\alpha)2^{2\alpha(l-1)}\left(8^{-(l-1)} +
    8^{-l}\right)N + 2(1+4^\alpha)2^{2\alpha(L-1)}8^{-(L-1)}N\\
  &= 2(1+4^\alpha)N\left(\sum_{l=1}^{L-1}
    \dfrac{9}{8}2^{2\alpha(l-1)}8^{-(l-1)} +
    2^{2\alpha(L-1)}8^{-(L-1)}\right) \\
  &\leq \dfrac{9(1+4^\alpha)}{4}N\sum_{l=1}^L 2^{(2\alpha-3)(l-1)}.
\end{align*}

If $\alpha < 3/2$, then the summation on the right-hand side
increases monotonically as $L$ increases, and converges to a finite
number as $L$ tends to infinity, with the limit depending on the
convergence rate $\alpha$ only. Thus, in this case, the computational
cost grows linearly as $N$ increases.
\begin{equation}
\label{eq:66}
  \C_{MLMC1} = O(N).
\end{equation}

If $\alpha = 3/2$, then
\begin{equation*}
  \C_{MLMC1} \leq \dfrac{9(1+4^\alpha)}{4} N L.
\end{equation*}
With $L$ as given in \eqref{eq:55}, we conclude that
\begin{equation}
  \label{eq:56}
  \C_{MLMC1} = O(N(\log\delta[U] + \log N)).
\end{equation}
We note that the case where $\alpha = 3/2$ {\it exactly} is rare in
practice. But the result obtained here, together with the result for
the $\alpha < 3/2$, indicates that, with larger $\alpha$, the
computational cost will increase faster as $N$ increases.

Finally, if $\alpha > 3/2$, then
\begin{equation*}
  \C_{MLMC1} = \dfrac{9(1+4^\alpha)}{4}N\cdot\dfrac{2^{(2\alpha - 3)L}
    - 1}{2^{2\alpha -3} - 1}.
\end{equation*}
Upon substituting the expression \eqref{eq:55} for $L$ in the above,
we obtain that
\begin{equation}
  \label{eq:57}
  \C_{MLMC1} = O\left( \delta[U]^\frac{2\alpha-3}{\alpha} \cdot N^{1+\frac{2\alpha-3}{3}}\right).
\end{equation}
In this case, the computational cost grows polynomially in $N$ and
$\delta[U]$, similar to the situation with the classical Monte Carlo
method (see
\eqref{eq:12}), but the exponents on both $\delta[U]$ and $N$ are
lower in the case here, indicating that, even if the convergence rate
$\alpha$ is greater than $3/2$, there still are potential savings in
computational time by choosing the MLMC method.

The problem with this strategy is that the error depends on the number
of levels, which may be large. In the following strategies, we amply
$M_l$ by certain factors so that the summation in \eqref{eq:20}
actually converges even as the number of levels goes to infinity, so
that the final error is actually independent of the number of levels
taken.\\

\noindent{\itshape Strategy \#2}\\
Under this strategy, we make the error term in the summation on the
right-hand side of \eqref{eq:20} decrease exponentially as the level
number $l$ goes up, that is, we set
\begin{equation*}
  \dfrac{\sqrt{2(1+4^\alpha)}\cdot r_l^\alpha}{\sqrt{M_l}} =
  \left(\dfrac{1}{2}\right)^{l-1} r_1^\alpha,
\end{equation*}
which leads to
\begin{equation}
  \label{eq:26}
  M_l = 2(1+4^\alpha)\cdot 4^{(l-1)(\alpha+1)}.
\end{equation}
To ensure that the error term due to the inherent variance of the
system be on the same level as the discretization error, we require
that
\begin{equation*}
  \dfrac{\delta[U]}{\sqrt{M_L}} = e,
\end{equation*}
from which we infer that
\begin{equation*}
  M_L =  \left(\dfrac{\delta[U]}{e}\right)^2.
\end{equation*}
Using the formula \eqref{eq:26} for $M_L$, we obtain a lower bound for
the number of levels required,
\begin{equation}
  \label{eq:27}
  L = \dfrac{2\log\delta[U] - 2\log e - \log
    2(1+4^\alpha)}{(\alpha+1)\log 4} + 1,
\end{equation}
or, using \eqref{eq:18},
\begin{equation}
  \label{eq:28}
  L = \dfrac{2\log\delta[U] - 2\alpha\log r_1 - 2\log C_2- \log
    2(1+4^\alpha)}{(\alpha+1)\log 4} + 1.
\end{equation}

We note that, from \eqref{eq:3},
\begin{equation}
  \label{eq:29}
  \log r_1 = \dfrac{\log C_1 - \log  N}{3}.
\end{equation}
Hence, we have
\begin{equation}
  \label{eq:30}
  L = \dfrac{2}{(\alpha+1)\log 4}\log\delta[U] +
  \dfrac{2\alpha}{3(\alpha + 1)\log 4}\log N + 1 
   -
  \dfrac{\frac{2\alpha}{3}\log C_1 + 2\log C_2 + \log
    2(1+4^\alpha)}{(\alpha+1)\log 4}.
\end{equation}
This expression indicates that, generally, large variance in the
analytic solution requires more levels. It is also clear from the
expression that, under the same convergence rate
$\alpha$, and the same constants $C_1$ for computational cost and
$C_2$ for the error, finer mesh (larger $N$) will also requires more
levels.

Under this strategy, the error in the $L$-level mean is independent of
the number of levels, for
\begin{equation}
\label{eq:32}
  \|\mbox{\rm\sffamily E}[U] - \mbox{\rm\sffamily E}^L[U]\|_\lop \leq e\left\{1 + \sum_{l=1}^L
    \left(\dfrac{1}{2}\right)^{l-1} + 1\right\} \leq 4e.
\end{equation}

We denote the computational cost under this strategy by
$\mathcal{C}_{MLMC2}$. It is calculated as follows,
\begin{align*}
  \mathcal{C}_{MLMC2} &= \sum_{l=1}^L \tilde M_l N_l
  = M_1 N_1 + \sum_{l=2}^L (M_{l-1}+M_l)N_l\\
  &= 2(1+4^\alpha)N + \sum_{l=2}^L 2(1+4^\alpha)\left(
   4^{(l-2)(\alpha+1)} + 4^{(l-1)(\alpha+1)}\right)\cdot 8^{-(l-1)}
 N\\
  & = 2(1+4^\alpha)(1+4^{-(\alpha+1)}) N \sum_{l=1}^L 2^{(l-1)(2\alpha -1)}.
\end{align*}

If $\alpha < 1/2$, then the summation on the right-hand side increases
monotonically as $L$ increases, and converges to a limit as $L$ tends
to infinity. The limit depends on the convergence rate $\alpha$ only.
Thus, in this case, the computational cost $\mathcal{C}_{MLMC2}$ grows linearly
in $N$. Specifically,
\begin{equation}
  \label{eq:58}
  \C_{MLMC2} \leq \dfrac{2(1+4^\alpha)(1+4^{-(\alpha+1)})}{1-2^{2\alpha
      - 1}} N.
\end{equation}
\begin{equation}
\label{eq:67}
  \C_{MLMC2} = O(N).
\end{equation}

If $\alpha = 1/2$, then
\begin{equation*}
  \C_{MLMC2} = 2(1+4^\alpha)(1+4^{-(\alpha+1)}) NL.
\end{equation*}
With $L$ as given in \eqref{eq:30}, we conclude that
\begin{equation}
  \label{eq:59}
   \C_{MLMC2} = O(N(\log\delta[U] + \log N)).
\end{equation}

We now consider the more common scenario where $\alpha > 1/2$.
\begin{equation}
  \label{eq:31}
  \mathcal{C}_{MLMC2} = 2(1+4^\alpha)(1-4^{-(\alpha + 1)}) N\cdot
  \dfrac{2^{(2\alpha -1)L}-1}{2^{2\alpha - 1} - 1}.
\end{equation}
Substitute the expression \eqref{eq:30} for $L$ into the above, we
obtain that
\begin{equation}
  \label{eq:33}
  \mathcal{C}_{MLMC2} = C_3 \cdot
  \dfrac{2(1+4^\alpha)\cdot(1-4^{-(\alpha+1)})}{2^{(2\alpha -1 )} -
    1}\delta[U]^{\frac{2\alpha -1}{\alpha+1}}\cdot
    N^{1+\frac{\alpha(2\alpha -1)}{3(\alpha +1)}}.
\end{equation}
\begin{equation}
  \label{eq:25}
  \C_{MLMC2} = O\left(\delta[U]^{\frac{2\alpha -1}{\alpha+1}}\cdot
    N^{1+\frac{\alpha(2\alpha -1)}{3(\alpha +1)}}\right).
\end{equation}
Comparing with the conventional MC method, the computational
cost for the MLMC method under the current strategy still grows
polynomially as $N$ increases, but at a lower degree, for it is
trivial to verify that
\begin{equation*}
  \dfrac{\alpha(2\alpha-1)}{3(\alpha+1)} \leq \dfrac{2}{3}.
\end{equation*}
The impact of the variance in the
analytical solution on the computational cost is also lower, for it is
obvious that
\begin{align*}
  \dfrac{2\alpha -1}{\alpha + 1} &\leq 2.
\end{align*}

\noindent{\itshape Strategy \#3}\\
This strategy chooses a sample size so that each term on the right-hand
side of \eqref{eq:20}, including the individual terms in the
summation, contributes equally to the total error, and then amplify
the sample size by a level dependent factor to ensure
convergence. With $\sigma>0$ being a positive parameter, we
let
\begin{equation}
  \label{eq:34}
  M_l = 2(1+4^\alpha)(L-l+1)^{2(1+\sigma)}\cdot 2^{2\alpha(l-1)}.
\end{equation}
As before, the number of levels is determined by requiring that the sample size
at the highest level satisfies the relation \eqref{eq:22}, which leads
to
\begin{equation}
  \label{eq:35}
  L = \dfrac{2\log\delta[U] - 2\log e - \log
    2(1+4^\alpha)}{2\alpha\log 2} + 1,
\end{equation}
or, using \eqref{eq:18},
\begin{equation}
  \label{eq:36}
  L = \dfrac{\log\delta[U]}{\alpha\log 2} - \dfrac{\log r_1}{\log
    2} +1 - \dfrac{2\log C_2 + \log 2(1+4^\alpha)}{2\alpha\log 2}.
\end{equation}
We note from \eqref{eq:3} that
\begin{equation*}
  \log r_1 = \dfrac{\log C_1 - \log N}{3}.
\end{equation*}
Substituting this expression into \eqref{eq:36} yields
\begin{equation}
\label{eq:38}
  L = \dfrac{\log\delta[U]}{\alpha\log 2} + \dfrac{\log N}{3\log
    2} +1 - \dfrac{\log C_1}{3\log 2} - \dfrac{2\log C_2 + \log
    2(1+4^\alpha)}{2\alpha\log 2}.
\end{equation}
The expression on the right-hand size indicates that, generally, a
larger variance in the analytical solution requires more levels. Under
the same order of convergence ($\alpha$), and the same constant
coefficients $C_1$ and $C_2$, a higher number of degrees of freedom ($N$,
or, in other words, a finer mesh) also requires more levels.

Under this strategy, the total error \eqref{eq:20} in the sample mean
can be estimated,
\begin{align*}
  &\|\mbox{\rm\sffamily E}[U] - \mbox{\rm\sffamily E}^L[U]\|_\lop
  \leq  e \left\{ 1 + \sum_{l=1}^L (L-l+1)^{-(1+\sigma)} +
      1\right\}
  =  e \left\{ 2 + \sum_{l=1}^L l^{-(1+\sigma)}\right\}.
\end{align*}
We note that, thanks to the positiveness of the parameter $\sigma$,
the summation converges even as $L$ tends to infinity. We can bound
the summation by an  integral, and we have
\begin{equation}
  \label{eq:39}
  \|\mbox{\rm\sffamily E}[U] - \mbox{\rm\sffamily E}^L[U]\|_\lop \leq  e \left(3 + \int_1^\infty
    l^{-(1+\sigma)} dl\right) \leq \left(3+\dfrac{1}{\sigma}\right)
    e.
\end{equation}
The computational cost $\mathcal{C}_{MLMC3}$ can also be estimated,
\begin{align*}
  \C_{MLMC3} =& \sum_{l=1}^L \tilde M_l N_l= \sum_{l=1}^L(M_{l-1} + M_l) N_l= \sum_{l=1}^{L-1} M_l(N_l + N_{l+1}) + M_L N_L\\
  =& 2(1+4^\alpha)N\left\{\sum_{l=1}^{L-1} (L-l +1)^{2(1+\sigma)}\cdot
    2^{(2\alpha-3)(l-1)}\cdot (1+2^{-3}) +
    2^{2\alpha-3)(L-1)}\right\}.
\end{align*}
We note that the last term in the curly bracket can be rolled over
into the summation, and an inequality follows,
\begin{equation}
\label{eq:40}
  \C_{MLMC3} \leq\dfrac{9(1+4^\alpha)}{4} N \sum_{l=1}^L (L-l+1)^{2(1+\sigma)}\cdot
  2^{(2\alpha -3)(l-1)}.
\end{equation}
If $\alpha < 3/2$, then
\begin{align*}
  \C_{MLMC3} \leq& \dfrac{9(1+4^\alpha)}{4} N L^{2(1+\sigma)}
  \sum_{l=1}^L 2^{2(\alpha-3)(l-1)}
  \leq& \dfrac{9(1+4^\alpha)}{4} N
  L^{2(1+\sigma)}\cdot\dfrac{1}{1-2^{2\alpha -3}}\\
  =& \dfrac{9(1+4^\alpha)}{4(1-2^{2\alpha -3})} N
  L^{2(1+\sigma)}.
\end{align*}
Substituting \eqref{eq:38} into the expression above, we find that
\begin{equation}
  \label{eq:41}
  \C_{MLMC3} \leq \dfrac{9(1+4^\alpha)}{4(1-2^{(2\alpha -3)})}\cdot
  N\cdot \left(\dfrac{\log\delta[U]}{\alpha\log 2} + \dfrac{\log
      N}{3\log 2} + C\right)^{2(1+\sigma)}.
\end{equation}
From the above, we conclude that
\begin{equation}
  \label{eq:42}
  \C_{MLMC3} \sim N\cdot (\log\delta[U]+\log N )^{2(1+\sigma)}.
\end{equation}

If $\alpha = 3/2$, then
\begin{align*}
  \C_{MLMC3} =& \dfrac{9(1+4^\alpha)}{4}\cdot N \cdot \sum_{l=1}^L
  l^{2(1+\sigma)}
  \leq \dfrac{9(1+4^\alpha)}{4} \cdot N \cdot \int_1^{L+1}
l^{2(1+\sigma)} dl\\
\leq & \dfrac{9(1+4^\alpha)}{4(2\sigma +3)} \cdot N \cdot \left\{
  (L+1)^{2\sigma +3} \right\}
= \dfrac{9(1+4^\alpha)}{4(2\sigma +3)} \cdot N \cdot \left(
  \dfrac{\log\delta[U]}{\alpha\log 2} + \dfrac{\log N}{3\log 2} +
  C\right)^{2\sigma + 3},
\end{align*}
where
\begin{equation*}
  C = 2 - \dfrac{\log C_1}{3\log 2} - \dfrac{2\log C_2 + \log
    2(1+4^\alpha)}{2\alpha\log 2}.
\end{equation*}
Therefore, for this case,
\begin{equation}
  \label{eq:43}
  \C_{MLMC3} \sim N\cdot(\log\delta[U] + \log N)^{2\sigma + 3}.
\end{equation}

If $\alpha > 3/2$, then the situation is more complicated.
\begin{align*}
  \C_{MLMC3} &\leq\dfrac{9(1+4^\alpha)}{4} N \sum_{l=1}^L
  (L-l+1)^{2(1+\sigma)}\cdot
  2^{(2\alpha -3)(l-1)}\\
  &\leq \dfrac{9(1+4^\alpha)}{4} N L^{2(1+\sigma)} \sum_{l=1}^L
  2^{(2\alpha -3)(l-1)}.
\end{align*}
Using the expression \eqref{eq:38}, we determine that
\begin{equation}
  \label{eq:60}
  \C_{MLMC3} = O\left((\log\delta[U] + \log N)^{2(1+\sigma)}\cdot
    \delta[U]^\frac{2\alpha -3}{\alpha}\cdot N^{1+\frac{2\alpha - 3}{3}}\right).
\end{equation}
This resembles the situation under Strategy \#2, and the cost grows
polynomially as $N$ increases.
\\

\noindent{\itshape Strategy \#4}\\
It is similar to Strategy \#3, but the sample size are amplified at
higher levels (lower resolutions). We set
\begin{equation}
  \label{eq:45}
  M_l = 2(1+4^\alpha)\cdot l^{2(1+\sigma)} \cdot 2^{2\alpha(l-1)}.
\end{equation}
To determine the number of levels $L$, we require $M_L$ to satisfy the
relation \eqref{eq:22},
\begin{align}
  2(1+4^\alpha)\cdot L^{2(1+\sigma)}\cdot 2^{2\alpha(L-1)}  = &
  \dfrac{\delta[U]^2}{e^2},\nonumber\\
L^{2(1+\sigma)}\cdot 2^{2\alpha(L-1)}  =&
  \dfrac{\delta[U]^2}{e^2\cdot 2(1+4^\alpha)}.\label{eq:46}
\end{align}
The number of levels cannot be solved for explicitly from
\eqref{eq:46}. But it is clear that it is smaller than that of
Strategy
\#3. This strategy leads to the same total error in the $L$-level
sample mean,
\begin{equation}
  \label{eq:47}
  \|\mbox{\rm\sffamily E}[U] - \mbox{\rm\sffamily E}^L[U]\|_\lop \leq
  \left(3 + \dfrac{1}{\sigma}\right) e.
\end{equation}

We denote the computational cost under this strategy by $\C_{MLMC4}$,
\begin{align*}
  \C_{MLMC4} =& \sum_{l=1}^L \tilde M_l \cdot N_l
  = \sum_{l=1}^{L-1} M_l\cdot (N_l + N_{l+1}) + M_L\cdot N_L\\
  \leq & \dfrac{9(1+4^\alpha)}{4}\cdot N\cdot \sum_{l=1}^L
  l^{2(1+\sigma)}\cdot 2^{(2\alpha-3)(l-1)}.
\end{align*}
If $\alpha < 3/2$, then the summation on the right-hand side
converges. The limit, denoted by $C_{\alpha,\sigma}$, depends on the
parameters $\alpha$ and $\sigma$ only. Thus we have the estimate
\begin{equation}
\label{eq:61}
  \C_{MLMC4} \leq \dfrac{9(1+4^\alpha)}{4} C_{\alpha,\sigma}N.
\end{equation}
\begin{equation}
  \label{eq:44}
  \C_{MLMC4}  = \mathcal{O}(N).
\end{equation}

If $\alpha = 3/2$, then
\begin{align*}
  \C_{MLMC4} &= \dfrac{9(1+4^\alpha)}{4}\cdot N\sum_{l=1}^L
  l^{2(1+\sigma)}
  \leq \dfrac{9(1+4^\alpha)}{4(2\sigma +3)}\cdot N\cdot
  (L+1)^{2\sigma + 3}.
\end{align*}
The cost is the same as $\C_{MLMC3}$ for the same value of $\alpha$.
\begin{equation}
\label{eq:65}
  \C_{MLMC4} =\mathcal{O}\left( N\cdot(\log\delta[U] + \log N)^{2\sigma + 3}\right).
\end{equation}

If $\alpha > 3/2$, then
\begin{equation}
  \label{eq:48}
  \C_{MLMC4} = \dfrac{9(1+4^\alpha)}{4} \cdot N\cdot \sum_{l=1}^L
  l^{2(1+\sigma)} \cdot 2^{(2\alpha-3)(l-1)}.
\end{equation}
In this case, $\C_{MLMC4}$ is greater than $\C_{MLMC3}$, but shares the
same estimate, that is,
\begin{equation}
\label{eq:62}
  \C_{MLMC4} = \mathcal{O}\left(N\cdot(\log\delta[U] + \log N)^{2(1+\sigma)}\cdot
    \delta[U]^\frac{2\alpha -3}{\alpha}\cdot N^\frac{2\alpha - 3}{3}\right).
\end{equation}

\begin{sidewaystable}[H]\small
  \centering
  \ra{1.3}
  \begin{tabular}{@{}c|cccccc@{}}\toprule
    { } & Linear & Quasilinear  & Polynomial \\
\midrule
Classical MC &  &  & $\mathcal{O}(\delta[U]^2 N^{1+\frac{2}{3}\alpha})$ \\
 & & & \\
 \midrule
Strategy \#1 & $\mathcal{O}(N)$ & $\mathcal{O}\left(N(\log\delta[U]+\log
  N)\right)$  & $\mathcal{O}\left(\delta[U]^\frac{2\alpha
    -3}{\alpha}\cdot N^{1+\frac{2\alpha-3}{3}}\right)$\\
 { } & $(\alpha <3/2)$ &  $(\alpha = 3/2)$ & $(\alpha > 3/2)$ \\
 \midrule
Strategy \#2 & $\mathcal{O}(N)$ & $\mathcal{O}\left(N(\log\delta[U]+\log
  N)\right)$  & $\mathcal{O}\left(\delta[U]^\frac{2\alpha
    -1}{\alpha+1}\cdot N^{1+\frac{\alpha(2\alpha-1)}{3(\alpha+1)}}\right)$\\
 { } & $(\alpha <1/2)$ &  $(\alpha = 1/2)$ & $(\alpha > 1/2)$ \\
 \midrule
Strategy \#3 & $\mathcal{O}\left(N\cdot(\log\delta[U] + \log
  N)^{2(1+\sigma)}\right)$  & $\mathcal{O}\left(N\cdot(\log\delta[U] + \log
  N)^{2\sigma+3}\right)$ &  $\mathcal{O}\left((\log\delta[U] + \log
  N)^{2(1+\sigma)}\cdot\delta[U]^\frac{2\alpha-3}{\alpha}\cdot N^{1+\frac{2\alpha-3}{3}}\right)$\\
 { } & $(\alpha <3/2)$ &  $(\alpha = 3/2)$ & $(\alpha > 3/2)$ \\
 \midrule
Strategy \#4 & $\mathcal{O}(N)$ & $\mathcal{O}\left(N(\log\delta[U]+\log
  N)^{2\sigma + 3}\right)$  & $\mathcal{O}\left((\log\delta[U] + \log
  N)^{2(1+\sigma)}\cdot\delta[U]^\frac{2\alpha-3}{\alpha}\cdot
  N^{1+\frac{2\alpha-3}{3}}\right)$ \\
 { } & $(\alpha <3/2)$ &  $(\alpha = 3/2)$ & $(\alpha > 3/2)$ \\
\bottomrule
  \end{tabular}

    \caption{Comparison of growth rates for the classical MC
    method and the MLMC method under different
    strategies. $\delta[U]$ represents the standard deviation in the
    true solution, $N$ the computational cost of an individual
    simulation at the highest resolution, $\alpha$ the convergence
    rate, and $\sigma$ an arbitrary positive parameter chosen by the
    user. }
  \label{tab:comparison}

\end{sidewaystable}

The computational cost for each strategy, as well as the cost for the
classical MC method, are summarized in Table \ref{tab:comparison}. All
strategies, except Strategy \#3, experience three stages of cost
growth, depending on the convergence rate $\alpha$: linear,
quasi-linear, and polynomial. When the convergence rate is high, the
computational cost for all strategies grow polynomially, similar to
the situation of the classical MC method. But the degrees of the
polynomials are lower, offering potential savings in computing
times. Strategy \#3 appears disadvantage in that it lacks linear
growth for the computational cost, apparently due to the fact that the
sample size at the lowest level (highest resolution) is amplified.

\subsection{Estimates}\label{sec:estimates}
Under each one of the strategies discussed above, the calculation of
the number of levels, the sample size at each level, the error in the
$L$-level sample mean, and the computational cost depend on a few key
parameters, namely $\delta[U]$, the standard deviation in the true
solution, $\alpha$, the convergence rate of the numerical scheme
regarding the QoI, and $e$, the
$L^2$-norm of the error in the first approximation $U_1$.  Determining the
true values of these parameters touches upon several fundamental
mathematical and numerical issues that, in many cases involving
real-world applications, are completely open. For example, for many
nonlinear systems, e.g.,~the three-dimensional Navier-Stokes equations
governing fluids, the existence and uniqueness of a global solution is
still an open question. Similarly, the numerical analysis to determine
the convergence rate of numerical schemes for nonlinear systems is
very challenging, even not possible. We leave these theoretical issues
to future endeavors. In the current work, we explore approaches to
estimate these parameters from the discrete simulation data.

The standard deviation $\delta[U]$ in the true solution can be
approximated by the unbiased sample variance (\cite{Kenney:1951ta}),
\begin{equation}
  \label{eq:37}
  \delta[U] \approx \dfrac{1}{M_l -1} \sum_{m=1}^{M_l} \left( U^m_l
    - \mbox{\rm\sffamily E}_{M_l}[U_l]\right)^2.
\end{equation}

The convergence rate $\alpha$ cannot be calculated directly using the
$L^2$-norm of the error in $U_l$ and  the
relation \eqref{eq:64}, since the true solution $U$ is not
available. Instead, we use the standard deviation of the difference
between solutions at two consecutive levels, i.e.~$\delta[U_l -
U_{l+1}]$. 
Instead
of the coefficient $\sqrt{2(1+4^\alpha)}C_2$ on the right-hand side of
\eqref{eq:49}, we assume that there exists another constant $C_3$ such
that
\begin{equation}
  \label{eq:50}
  \delta[U_l - U_{l+1}] = C_3 r_l^\alpha.
\end{equation}
The computation of $\alpha$ will not be affected by the value of
$C_3$, since
\begin{equation}
\label{eq:63}
  \dfrac{\delta[U_1 - U_2]}{\delta[U_2 - U_3]} =
  \left(\dfrac{r_1}{r_2}\right)^\alpha = \left(\dfrac{1}{2}\right)^\alpha.
\end{equation}
Of course, in actual calculations, the standard deviation on the
left-hand side of \eqref{eq:50} will be replaced by the square root of
the unbiased sample variance (formula \eqref{eq:37}).

The $L^2$-norm of the error in the first approximation
$U_1$, $e$, cannot be calculated directly from \eqref{eq:64} either,
due to
the lack of the true solution $U$. 
Instead, using \eqref{eq:49}, and the convergence rate just computed,
we can obtain an estimate on $e$,
\begin{equation}
  \label{eq:51}
 e = \dfrac{\delta[U_1-U_2]}{\sqrt{2(1+4^\alpha)}}.
\end{equation}

\section{Numerical experiments using ACC}\label{sec:numer-exper-using}
The Antarctic Circumpolar Current (ACC) is a circular current
surrounding the Antarctic continent. It is the primary channel through
which the world's oceans (Atlantic, Indian, and Pacific)
communicate. Thanks to the predominant westerly wind in that region,
the current flows from west to east. The ACC is the strongest current
in the world, volume-wise. It is estimated that the volume transport is
about 135 Sv (1 Sv = $10^6$ $\mathrm{m}^3\,\mathrm{s}^{-1}$) through
the Drake passage
(\cite{Gent:2001fg,{Hughes:1999iu},{Warren:1996cx}}), which is about
135 times the total volume transport of all the rivers in the
world. The above estimate is a time average; the actual volume
transport oscillate on seasonal and intradecadal scales.

\begin{figure}[h]
  \centerline{\includegraphics[width=4in]{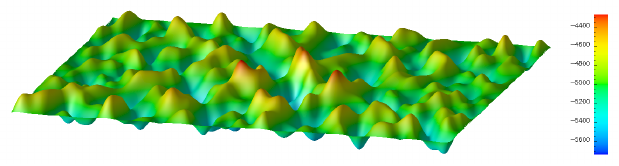}}
  \caption{The random bottom topography sample \# 1.}
  \label{fig:topo-sample}
\end{figure}

Here, we demonstrate how the MLMC method can be
combined with an ocean circulation model to quantify the volume
transport of the ACC. In order to stay focused on the methodology that
is being explored here, we sharply reduce the physics of this problem
while still retain its essential features. The fluid domain is a
re-entrant rectangle that is $2,000$ km long and $1,733$ km wide; the same size
as \cite{Chen:2016cr}, and also see \cite{McWilliams:1981up}.
The flow
is governed by a three-layer isopycnal model, which reads
\begin{equation}
\label{eq:52}
  \left\{\begin{aligned}
     & \dfrac{\p h_i}{\p t} + \nabla\cdot\left(h_i \ub_i\right) = 0,\\
     & \dfrac{\p\ub_i}{\p t} + h_iq_i\kb\times\ub_i =
     -\nabla\left(\dfrac{\phi_i}{\rho_0}+K_i\right) + \mathbf{D}_i + \mathbf{F}_i, \\
     & \dfrac{\p}{\p t} (h_i\sigma_i) + \nabla\cdot\left(h_i\sigma_i \ub_i\right) = 0,\\
    \end{aligned}\right.
\end{equation}
where $i=1,2,3$ is the layer index starting at the ocean surface. The prognostic variables $h_i$,
$\ub_i$ and $\sigma_i$ denote the layer thickness, horizontal
velocity, and some tracer
respectively, and the diagnostic variables $q_i$, $\phi_i$ and $K_i$
denote the potential vorticity, Montgomery potential and kinetic
energy, respectively, and they are defined as
\begin{align*}
  & q_i = \dfrac{\nabla\times\ub_i + f}{h_i},\qquad i=1,2,3, \\
  & K_i = \frac{1}{2}|\ub_i|^2,\qquad i=1,2,3,\\
  & \phi_1 = p_0 + \rho_1g(h_1+h_2+h_3+b),\\
  & \phi_2 = \phi_1 + (\rho_2-\rho_1)g(h_2+h_3+b),\\
  & \phi_3 = \phi_2 + (\rho_3-\rho_2)g(h_3+b),
\end{align*}
with $p_0$ denoting the surface pressure and $b$ the bathymetry;
$\mathbf{D}_i$ denotes the horizontal viscous diffusion, which usually takes the form
of harmonic or biharmonic diffusion.
The external forcing term $\mathbf{F}_i$ for
each layer is specified as follows,
\begin{equation}
\label{eq:53}
  \mathbf{F}_i = \left\{
    \begin{aligned}
      &\dfrac{\bs{\tau}}{\rho_1 h_1} \textrm{ (wind stress) }, & &i = 1,\\
      & 0, & &i =2,\\
      &-\mathbf{d}\textrm{ (bottom drag) }, & & i=3.
    \end{aligned}\right.
\end{equation}
The model is made up of
three isopycnal layers with mean layer thickness of 500 m, 1250 m and
3250 m and with densities of $1010\, \kg\,\m^{-3}$, $1013\, \kg\,\m^{-3}$
and $1016\,\kg\,\m^{-3}$. The system is forced by a zonal wind stress
on the top layer with the form
$$\tau = \tau_0\sin^2 \left(\dfrac{\pi y}{L_y}\right),$$
where $\tau_0 = 0.1\textrm{N}\,\m^{-2}$. The uncertainty in the model
is presented by the bottom topography. We assume that the bottom
topography of the domain is largely flat with small but random
features,
\begin{equation*}
  b = \sum_{k,l = 4}^{20} \dfrac{H}{k^2 + l^2}\left(a_{kl}(\omega)
    \cos\left(\dfrac{2\pi k x}{L_x}\right) +
    b_{kl}(\omega)\sin\left(\dfrac{2\pi k
        x}{L_x}\right)\right)\sin\left(\dfrac{l\pi y}{Ly}\right),
\end{equation*}
where $a_{kl}(\omega)$ and $b_{kl}(\omega)$ are random variables. Thus the
bottom is controlled by 578 random parameters. One sample of the
topography is shown in Figure \ref{fig:topo-sample}.
Similar types of bottom
topography profiles have been used by \cite{TREGUIER:1990wc}.
The numerical simulations are conducted using the MPAS isopycnal ocean
model (\cite{Ringler:2013vw}). MPAS, which stands for Model Prediction
Across Scales, implements a C-grid finite difference / finite volume
scheme that is detailed in \cite{Thuburn:2009tb,
  {Ringler:2010io}}. MPAS utilizes arbitrarily unstructured
Delaunay-Voronoi tessellations (\cite{Du:1999gs, {Du:2003ka}}). For
this experiment, we have four levels of resolutions available: 10 km,
20 km, 40 km, and 80 km. To account for the effect of the unresolved
eddies, the biharmonic hyperviscosity is used. The viscosity
parameters are chosen to minimize the diffusive effect while still
ensure a stable simulation. For the aforementioned resolutions, the
viscosity parameters are $10^9\,\mathrm{m}^4\mathrm{s}^{-1}$,
$10^{10}\,\mathrm{m}^4\mathrm{s}^{-1}$,
$10^{11}\,\mathrm{m}^4\mathrm{s}^{-1}$,
and $10^{12}\,\mathrm{m}^4\mathrm{s}^{-1}$, respectively. At the
coarsest resolution (80 km), the Gent-McWilliams closure
(\cite{Gent:1990jt,{Gent:1995fa}}) is turned on,
with a constant parameter $400\,\mathrm{m}^2\mathrm{s}^{-1}$, to
account for the cross-channel transport and to prevent the top fluid layer
thickness from thinning to zero. GM is not used in any other higher
resolution simulations. The configurations for each mesh resolution
are summarized in Table \ref{tab:resolutions}.
Each simulation is run for 40 years to spin up the current. The output
data are saved every 10 days for the next 10 years.

\begin{table}[h]\small
  \centering
  \begin{tabular}[h]{@{}c|cccc@{}}\toprule
    { } & Eddy closures & Spatial DOFs & Time step (s) & Processes\\
\midrule
    10 km & Hyperviscosity  & 480,000 & 45 & 64 \\
    \midrule
    20 km & Hyperviscosity & 120,000 & 90 & 16 \\
    \midrule
    40 km & Hyperviscosity & 30,000 &  180  & 4 \\
    \midrule
    80 km & Hyper. + GM & 7,500 & 360 & 1 \\
\bottomrule
  \end{tabular}
  \caption{The configurations for each resolution. The spatial degrees
  of freedom (DOFs) is calculated as (number of cells + number of
  edges) $\times$ number of layers.}
  \label{tab:resolutions}
\end{table}

\begin{figure}[h]
  \centering
  \includegraphics[width=4.5in]{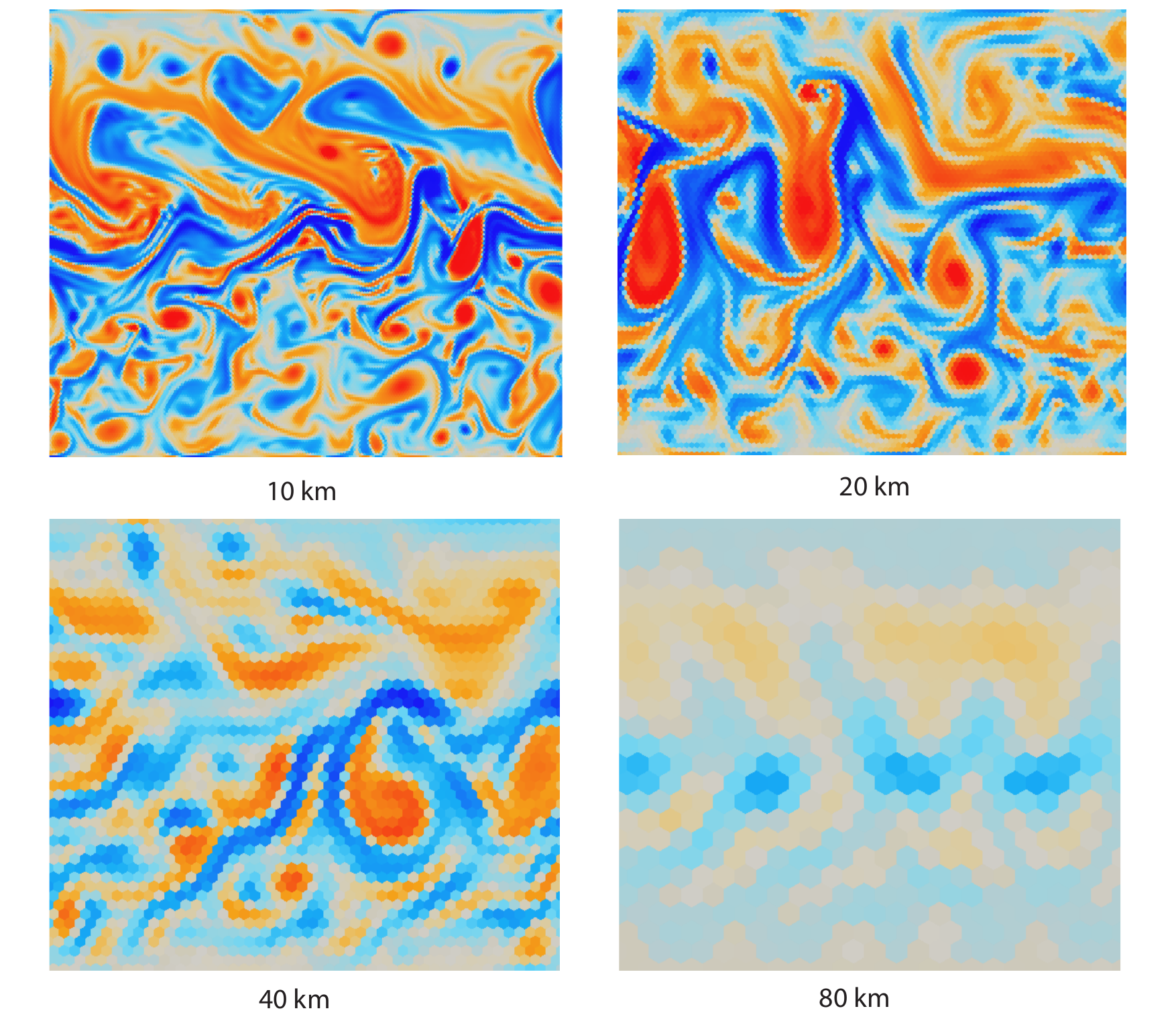}
  \caption{The snapshots of the vorticity field at year 40, 
  computed with the random bottom topography sample \#1.}
  \label{fig:vorticity-y40}
\end{figure}

\begin{figure}[h]
  \centering
  \includegraphics[width=4in]{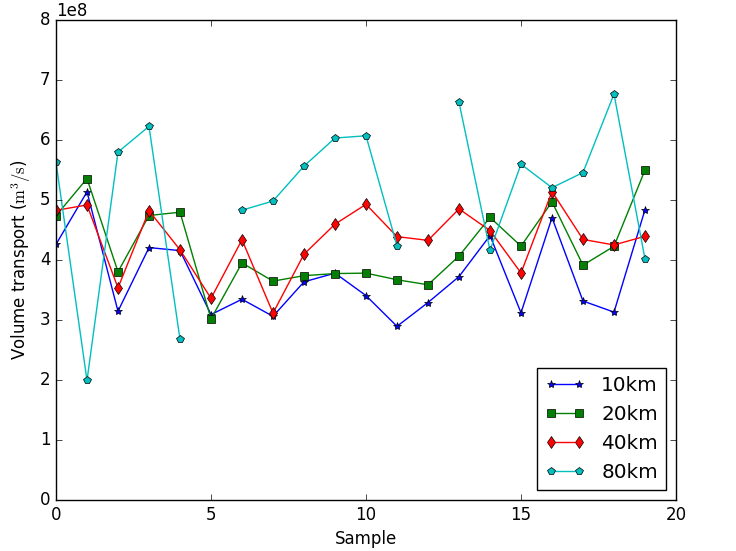}
  \caption{The changes of volume transport across a subset of the sample space.}
  \label{fig:trend}
\end{figure}

The interior of large-scale geophysical flows has Reynolds numbers on
the order of $10^{20}$. Thus the large-scale geophysical flows are
turbulent in nature, and mesoscale and submesoscale eddy activities
are important part of the ocean dynamics
(\cite{{McWilliams:1985fu},{Danabasoglu:1994uh},{Levy:2001dx},
  {Holland:2010jk}, {Brannigan:2015hz}}). For turbulent flows, the
pointwise instantaneous behavior of the flow is not
reliable anymore. But one can hope that observing the flow long enough
can reveal reliable and useful statistics about the flow. Figure
\ref{fig:vorticity-y40} shows the snapshots of the relative vorticity
field on Year 40 for mesh resolutions with the same bottom topography
profile. The highest resolution, 10km (Panel (a)), depicts a scene of
rapid mixing
by a wide range of mesoscale and submesoscale eddies.
As the mesh gets coarser, the level of eddy activities decrease. The
comparison also makes it clear that these flows are largely
independent of each other, for there appears to be no correlation
between the basic flow patterns of these simulations, other than the
fact that they are all west-to-east flows driven by a common
windstress.
However, a comparison of the volume transport by these simulations
over a common set of topographic profiles tells a different and
reassuring story. In Figure (\ref{fig:trend}), each curve
represents results on one mesh resolution. While on any particular
topography profile, the results from different resolutions do not agree,
the curves across all the 20 samples, especially those
for the 10 km, 20 km, and 40 km, largely follow the same pattern.
The agreement of the patterns of the curves
indicates that a great deal of information in the curve for the
highest resolutions is actually available in the curves of the lower
resolutions, and this is a vindication for the multi-level method
that we are pursuing here.

\begin{figure}[h]
  \centering
  \includegraphics[width=5in]{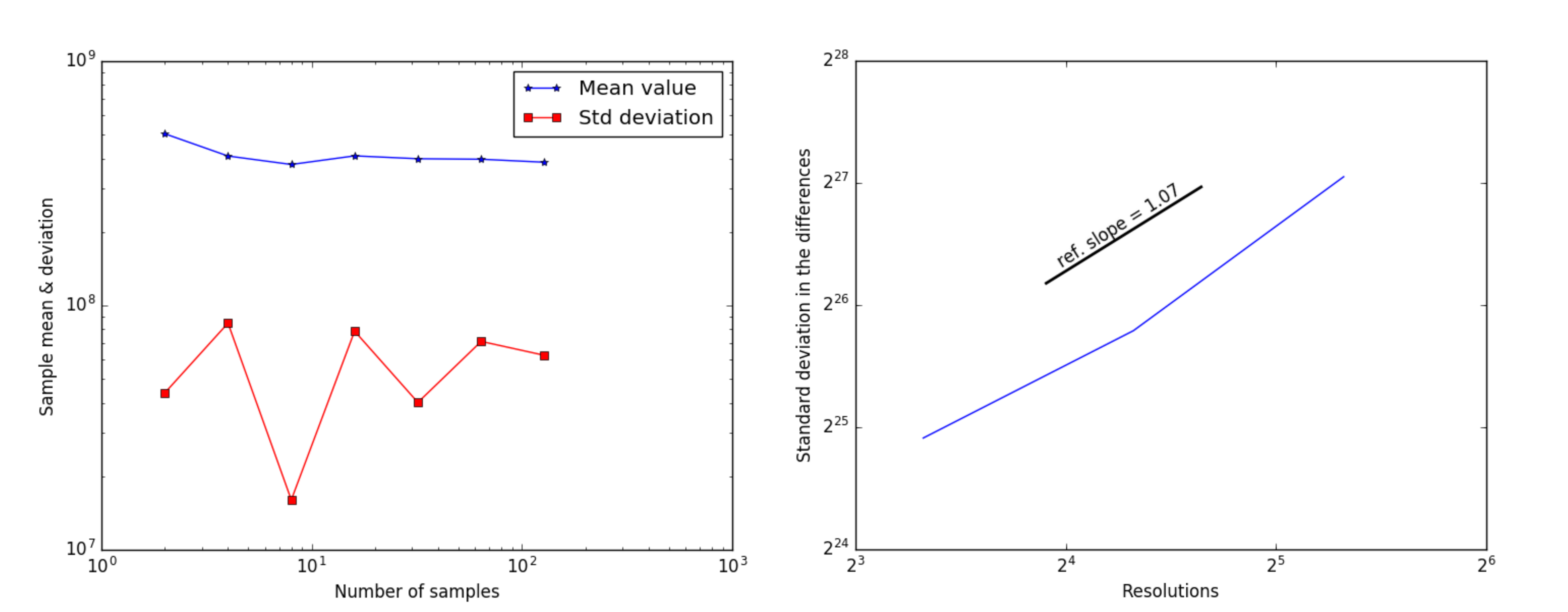}
  \caption{Sample mean and variance and the convergence rate. The mean
  and variance are estimated using samples from the 20km simulations.}
  \label{fig:rate}
\end{figure}

In order to set up the MLMC simulations under the
various strategies proposed before, three key parameters are needed:
the standard deviation $\delta[U]$ in the true solution, the error
$e$ in the
finest solutions, and the convergence rate $\alpha$. To fully
determine these key parameters requires the true solution itself $U$,
which is not available in any practical applications. But they can
easily estimated (see Section \ref{sec:estimates}). Using data from
the 20 km simulations, we compute the MC mean and the standard
deviation according to the formulae \eqref{eq:5} and \eqref{eq:37}. To
probe the sensitivity of these estimates to the sample sizes, we
compute the quantities with several independent sample sets with
varying sizes, and the results are shown in Figure \ref{fig:rate}
(left panel). Based on this figure, we take
\begin{equation*}
  \delta[U] \approx 7.36\times 10^7\textrm{ m}^3/\textrm{s}.
\end{equation*}
Using the formula \eqref{eq:63} and data from 10 km, 20 km, 40 km, and
80 km simulations (Figure \ref{fig:rate} (right panel)), the
convergence rate $\alpha$ is estimated to be
\begin{equation*}
  \alpha \approx 1.07.
\end{equation*}
This convergence appears slow but expected for long term simulations of
turbulent flows. The underlying numerical scheme, namely a C-grid
finite volume scheme, has been found to be accurate of orders $1\sim
2$ for laminar flows (\cite{Ringler:2010io}, also see
\cite{Chen:2013bl}). Finally the error $e$ in the finest solutions
is calculated using the formula \eqref{eq:51} and data from the 10km
and 20km simulations,
\begin{equation*}
  e \approx 9.60\times 10^6\textrm{ m}^3/\textrm{s}.
\end{equation*}

\begin{table}[h]\small
  \centering
  \ra{1.3}
  \begin{tabular}{@{}c|ccccc|c@{}}\toprule
    { } & $L$ & $m_1$ & $m_2$ & $m_3$ & $m_4$ & Comp.~load \\
\midrule
Classical MC & n/a & 59 & & & & 59 \\
\midrule
Strategy \#1 & 3 & 11 & 48 & 210 & & 22.4\\
\midrule
Strategy \#2 & 2 & 11 & 191 & & & 36.3\\
\midrule
Strategy \#3 & 3 & 876 & 763 & 210  & & 1096.1\\
\midrule
Strategy \#4 & 2 & 11 & 763 & & & 107\\
\bottomrule
  \end{tabular}
  \caption{Multi-level setup under different strategies}
  \label{tab:setup}
\end{table}

Using these estimated parameters and the formulae set forth under
various strategies propose in the previous section, we calculate the
number of levels, the sample size at each level, and the computational
load for each strategy for the multi-level method. The sample size and
computational load for
the classical Monte Carlo method are also calculated. The
computational load are calculated in terms of the the computational
load for one single simulation at the highest resolution (lowest
level). The issue of efficiency, overhead, etc.~are neglected. For
example, the classical MC method requires 59 simulations at
the highest resolution, and therefore its computational load is
59. The results are listed in Table \ref{tab:setup}. Several striking
features are present in the results. First of all, under all
strategies, the numbers of required levels are low (2 or 3). This can
be attributed to the fact that the error in the finest solutions are
high compare to the variance in the true solution. Second, the sample
sizes at the lowest level (highest resolution) are identical for
Strategy \#1, 2 and 4 (11 for all three). This is no coincidence. A
careful examination
of the formulae \eqref{eq:21}, \eqref{eq:26}, \eqref{eq:45} reveal
that, at the lowest level $l=1$, the sample sizes for Strategies \#1,
2, and 4 are identical, and depend on the convergence rate $\alpha$
only. Thus, irregardless of the actual highest resolution used, the
sample sizes for this model at the lowest level will remain the same
(=11) and identical for all three strategies.  Finally, for Strategy
\#3, the sample size  at the highest resolution is too high, and
results in a computational load even higher than that of the classical
Monte Carlo method. The reason is that this strategy requires a error
distribution that is low at the highest resolution, and high at the
lowest resolution. In terms of computational loads, Strategy \#1 is
the optimal choice, and it is followed by Strategy \#2. Both
strategies are better than the classical Monte Carlo method. Strategy
\#4 is actually more costly than the classical method, due to the
bloated sample size at the next level.

\begin{table}[h]\footnotesize
  \centering
  \ra{1.3}
  \begin{tabular}{@{}c|cccccc@{}}\toprule
    { } & Est.~volume & Est.~error & CPU time & CPU efficiency \\
    { } & transport ($\mathrm{m}^3/\mathrm{s}$) &
    ($\mathrm{m}^3/\mathrm{s}$) &(hours) & (DOFs
/ CPU second)\\
\midrule
Classical MC & $3.58\times 10^8$ & $1.92\times 10^7$ & 158,446 &
343,172 \\\midrule
Strategy \#1 & $3.55\times 10^8$ & $4.80\times 10^7$ & 45,917 &
449,719\\\midrule
Strategy \#2 & $3.57\times 10^8$  & $3.84\times 10^7$  & 68,457
&488,013\\
\bottomrule
  \end{tabular}
  \caption{Comparison between the classical MC, and the MMC under
    strategies \#1 and \#2. The efficiency is calculated as Total DOFs
    / Total CPU time. Total DOFs is calculated as Spatial DOFs
  $\times$ Time steps $\times$ Number of samples.}
  \label{tab:results}
\end{table}

We proceed to calculate the estimates and errors in the estimates using
the classical MC method and the MLMC method under strategies \#1 and 2. The results from the classical MC
can serve as a reference, since by the analysis of
\ref{sec:monte-carlo-method}, its error should be the
smallest. Strategies \#3 and 4 are not used
due to the shear sizes of their computational loads. The classical MC
and both Strategies \#1 and 2 produce similar estimates, $\sim
3.58\times 10^8\textrm{ m}^3/\mathrm{s}$, for the volume
transport (second column of Table \ref{tab:results}). The error for
each method is listed in the third column. The result of the classical
MC method has an error of about 5.4\%. The
error for Strategy \#1 is the largest, about 13.5\%. This is expected,
because the error for this strategy depends on the number of levels
(see \eqref{eq:24}), which is higher than that for Strategy \#2.

The primary advantage of MLMC is efficiency. The first
indicator for efficiency is of course the computational load that each
method will incur, which has already be listed in Table
\ref{tab:setup}. These numbers are the theoretical computational load,
and takes no consideration of computational overhead, parallelization,
etc. Here, we examine the actual efficiency for each
method. First, we look at
the total CPU hours used by each method (fourth column of Table
\ref{tab:results}).  The MLMC with strategy \#1 uses the least amount
of CPU time, 45,917 CPU hours, a saving of $71\%$ compared with the MC
method. Strategy \#2 use 68,457 CPU
hours, a saving of $67\%$. 

The savings of MLMC strategies in the actual CPU times are largely in
line with the savings in computational loads (Table \ref{tab:setup}),
but appear more dramatic than what the latter would suggest. This is
due to the increased CPU efficiency under the MLMC methods.
It is well known that MC methods are easy to parallelize,
and therefore highly scalable on supercomputers. The MLMC has the
potential to increase the efficiency over the classical MC even
further, by running more small-sized simulations and fewer large
simulations. Due to the large sizes of the
computations in this project, we are not able to perform a actual
scalability analysis, which involves running the experiment with
different numbers of total available processes. However, we can
indirectly examine the issue of scalability by comparing the efficiency for each CPU
core for the methods considered here (last column of Table
\ref{tab:results}).  Compared with the classical MC, Strategy \#1
increases the CPU efficiency by 31\%, and Strategy \#2 increases the
CPU efficiency even more, by 42\%.

\section{Discussions}\label{sec:discussions}
The success of the MLMC method relies on a crucial assumption, namely
that, a lot of the information contained in high-resolution
simulations is also available from low-resolution simulations, under
the identical or similar model configurations. The higher the
correlation, the better the MLMC method will work. For steady-state or
laminar flows, especially when the simulations are backed up by
rigorous error estimates, the correlation between high-resolution and
low-resolution solutions is high and quantifiable, and the MLMC method
works very well (see references cited in Introduction). For long-term
simulations of turbulent flows, the situation is different. It is
known that the pointwise behaviors of high- and low-resolution
solutions of turbulent flows are uncorrelated (Figure
\ref{fig:vorticity-y40}). But pointwise behaviors of turbulent flows
are of little interest. What are important are certain aggregated
quantities such as mean SST. Then, naturally arise
the questions as to whether these aggregated quantities are correlated
across different resolutions, and whether the MLMC method can be used
to save computation times. Through an experiment with the Antarctic
Circumpolar Current, the present work gives affirmative answers to both
of these questions. The conclusions drawn in this work cannot be
generalized universally to all turbulent flows, because, after all,
there is no universal theory for turbulent flows yet. But it is
reasonable to expect that the same results should hold in similar
situations. Specifically, the MLMC method can be effective in saving
computation times when the QoI demonstrates a certain
level of correlation across different resolutions.

Another objective of this paper is to explore how the MLMC simulation
can be set up. Four different strategies are presented, based on the
desired error distributions. One surprising finding is that the
performance of each strategy, with regard to computational cost,
depends on the convergence rate. For all strategies discussed, the
higher the convergence rate is, the faster the computational cost will
grow. This sounds counter-intuitive. Here, the focus is on how fast
the total computational cost will grow in terms of the computational
cost of a single high-resolution simulation (linearly, quadratically,
etc.) Of course, for the same highest resolution, a higher convergence
rate will eventually leads to more accurate results, and the
associated higher computational cost is a price paid for this higher
accuracy.

Among all the four strategies discussed in this work, Strategy \#3,
which amplify the sample sizes at high resolutions, seems to be of
little use, because of the unreasonably high cost. Strategy \#1 is the
most natural choice, but it may lead to bigger margin of errors if the
number of levels is high. In that situation, Strategy \#2 \& \#4 can
be used. In our experiment with the ACC, the highest resolution has
40,000 grid points, and Strategy \#2 outperforms Strategy \#4 with a
lower computational cost. But it should be kept in mind that, even at a
very modest convergence rate ($\alpha < 1/2$), the
computational cost for Strategy \#2 grow polynomially with respect to
the computational cost for a single high-resolution
simulation. Therefore, it is conceivable that, as higher resolution
are taken into use, Strategy \#4 will eventually outperform Strategy
\#2.

\section*{Acknowledgment}
This work was in part supported by  Simons
  Foundation (\#319070 to Qingshan Chen) and NSF of China (\#91330104 to Ju Ming).

\bibliographystyle{amsplain}
 \bibliography{references}

\end{document}